# Enhancement of spin-orbit coupling and magnetic scattering in hydrogenated graphene

Shimin Cao[1,2], Chuanwu Cao[1*], Shibing Tian[3], Jian-Hao Chen[1,2,4,5†]

[1]*International Center of Quantum Material, School of Physics, Peking University, Beijing 100871, China*
[2]*Beijing Academy of Quantum Information Sciences, Beijing 100193, China*
[3]*Institute of Physics, Chinese Academy of Sciences, Beijing 100190, China*
[4]*Key Laboratory for the Physics and Chemistry of Nanodevices, Peking University, Beijing 100871, China*
[5]*Interdisciplinary Institute of Light-Element Quantum Materials and Research Center for Light-Element Advanced Materials, Peking University, Beijing 100871, China*

**ABSTRACT**

Spin-orbit coupling can provide essential tools to manipulate electron spins in two dimensional materials like graphene, which is of great interest for both fundamental physics and spintronics application. In this work, we report the low field magneto-transport of *in-situ* hydrogenated graphene where hydrogen atoms are attached to the graphene surface in continuous low temperature and vacuum environment. Transition from weak localization to weak anti-localization with increasing hydrogen adatom density is observed, indicating enhancing Bychkov-Rashba type spin-orbit coupling in a mirror symmetry broken system. From the low temperature saturation of phase breaking scattering rate, the existence of spin-flip scattering is identified, which corroborates the existence of magnetic moments in hydrogenated graphene.

## I. INTRODUCTION

Spin-orbit coupling (SOC) is a relativistic effect which describes the mixing of electron spin and orbital motion. This process provides tools to the generation[1-3] and manipulation[4, 5] of spin polarization in electron systems, which coincide with the main goal of spintronics. Graphene, a two-dimensional atomic layer of carbon, has the advantage of high stability and accessibility[6, 7], high mobility and gate-tunability[8], as well as long spin diffusion length[9, 10], which makes it a suitable host for future spintronic applications. However, SOC in pristine graphene is expected to be weak due to relatively low electron velocity in carbon atoms[11]; furthermore, the first order of the atomic spin orbit coupling vanishes due to the honeycomb arrangement of the carbon atoms[12]. Therefore, introducing SOC into graphene could provide additional way to generate and manipulate spin, which has attracted extensive attention.

Hydrogenated graphene, e.g., graphene with hydrogen atoms covalently bonded to its surface, has been studied in a number of experiments recently[3, 13-18]. Hydrogenated graphene is a complex system due to the randomness of the location of the hydrogen adatoms, therefore has rich physics. Great enhancement of the SOC up to three orders of magnitude and strong non-local transport up to room temperature due to spin Hall effect are reported[3]. Large negative magnetoresistance (*MR*) is frequently observed in hydrogenated graphene, which is found to be a hint of magnetic coupling between hydrogen adatoms[13, 14]. Furthermore, hydrogen adatoms on bilayer graphene are found to energetically favor non-dimer sublattice over dimer sublattice, which is highly desirable for generating ferromagnetism[15].

---

[*] Currently in DQMP, University of Geneva.
[†] chenjianhao@pku.edu.cn

In this work, we report low field (B < 0.5 T) *MR* of *in-situ* hydrogenated graphene. We observed the transition between weak localization (WL) and weak anti-localization (WAL) which is tunable by hydrogen density and carrier density. The SOC strength extracted from the *MR* curves is found to be enhanced at high hydrogen density and low carrier density. Furthermore, the phase breaking rate exhibit a saturation at low temperature, which is a strong evidence in line with the scenario that hydrogen on graphene induced magnetic moments[19, 20].

## II. RESULT AND DISCUSSION

*In-situ* doping is a powerful tool to introduce novel interactions to two-dimensional materials. Various types of molecules, atoms or ions can be deposited or bombarded on the sample surface to introduce different types of interactions, so that the strength of interactions can be almost continuously controlled by the amount of doping and be detected via quantum transport, which is a great advantage comparing to other methods of introducing interactions like *ex-situ* doping[21, 22] or proximity effect[23, 24] in heterostructures of graphene and transition metal ditellurides (TMDs), albeit the proximity effect has the potential to introduce ordered interactions.

In this experiment, the hydrogen atoms are generated from a radio frequency atom source, flow towards the sample with low kinetic energy and covalently bond to the graphene surface[13]. The graphene sample is kept in ultra-high vacuum (UHV) environment of up to $10^{-9}$ torr to reduce unwanted impurity absorption except during the doping process, when the graphene sample is exposed to a partial pressure of $10^{-2}$ torr of hydrogen. The amount of hydrogen atoms on graphene can be controlled by the duration of the doping process (usually 5-10 seconds for each round of doping, 13 rounds of doping in this experiment) and quantitatively estimated from the conductance dependence on the carrier density of the sample after each hydrogenation round[13].

Figure1 shows the resistance of (a) pristine ($N_H = 0$) and (b) hydrogenated ($N_H = 7$) graphene, respectively. Before hydrogenation, the resistance of graphene is smaller than the resistance quantum $R_Q = h/e^2$ with weak temperature dependence. Mobility of pristine graphene is found to be 4000 cm$^2$/Vs, which is a fair value for graphene on SiO$_2$ substrate. After hydrogenation, the Dirac point shifts towards the negative gate voltages, which means that hydrogen adatoms electron doped graphene, consistent with previous angle-resolved photo emission spectroscopy study[16]. Furthermore, with large hydrogen concentration, the resistance of hydrogenated graphene can be larger than $R_Q$ near the Dirac point[13]. The resistivity of hydrogenated graphene is sensitive to temperature, but neither thermal activation model nor variable-range hopping model can be used to describe the resistivity vs. temperature data even at the Dirac point (see Supplementary Information figure S1).

Next, we focus on the low field ($B < 0.5$T) magneto-transport properties of hydrogenated graphene. During our experiment, $n_H$ increases monotonically from each round of hydrogenation. WL in graphene becomes apparent with a minimal number of $n_H$ and the WL signal increases with increasing $n_H$ (see Supplementary Information figure S2), which is not surprising and can be understood as the enhancement of WL signal due to increasing intervalley scattering from the hydrogen adatoms. What's more interesting is the evolution of the weak field *MR* at higher $n_H$ (from doping Round #6 to #11), which shows the systematic changes as a function of electron density ($n_e$) and $n_H$. Figure 2 shows *MR* of hydrogenated graphene at nine different combinations of $n_e$ and $n_H$ and at different temperatures. It can be seen that in the high $n_e$ regime (e.g. the difference between gate voltage and Dirac point gate voltage $V_g$ - $V_D \geq 40$ V), increasing $n_H$ simply suppresses the WL signal, as shown in Figure 2 (a-f); in the low $n_e$ regime (e.g. $V_g$ - $V_D \leq 10$ V), on the other hand, increasing $n_H$ causes a transition from WL to WAL. As will be shown in the discussion below, this is the first observation of tunable WL to WAL transition in graphene where WAL is cause by factors other than the π Berry phase in the material. With increasing temperature, both WL and WAL signal decay rapidly, consistent with the dependence of such effects on phase coherent length of the electrons, which in itself has strong temperature dependence.

Quantitatively, WL/WAL in graphene is caused by the interplay of various scattering mechanisms, usually including phase breaking scattering rate $\tau_\phi^{-1}$, intervalley scattering rate $\tau_i^{-1}$ and intravalley symmetry breaking scattering rate $\tau_*^{-1}$[25]. For hydrogenated graphene, since spin-orbit effect is known to be enhanced[3], it will also include Bychkov-Rashba (BR) type SOC scattering rate $\tau_{BR}^{-1}$ or Kane-Mele (KM) type SOC scattering rate $\tau_{KM}^{-1}$. Thus WL/WAL of graphene can be expressed as[25]:

$$\frac{2\pi h}{e^2}\frac{\Delta\rho}{\rho^2} = F\left(\frac{B}{B_\phi}\right) - F\left(\frac{B}{B_\phi+2B_i}\right) - 2F\left(\frac{B}{B_\phi+B_*}\right) - 2F\left(\frac{B}{B_\phi+B_{BR}+B_{KM}}\right) + 2F\left(\frac{B}{B_\phi+2B_i+B_{BR}+B_{KM}}\right) +$$
$$4F\left(\frac{B}{B_\phi+B_*+B_{BR}+B_{KM}}\right) - F\left(\frac{B}{B_\phi+2B_{BR}}\right) + F\left(\frac{B}{B_\phi+B_i+2B_{BR}}\right) + 2F\left(\frac{B}{B_\phi+2B_*+2B_{BR}}\right) \quad \text{Eq.(1)}$$

where $F(z) = ln(z) + \psi\left(\frac{1}{2}+\frac{1}{z}\right)$, $B_x = \frac{\hbar}{4De\tau_x} = \frac{\hbar}{4eL_x^2}$, $x = \phi, i, *, BR, KM$, $\psi$ represents digamma function, $D = v_F L_{mfp}/2$ is the diffusion constant determined by Fermi velocity $v_F = 10^6 m/s$ and mean free path $L_{mfp} = (\sigma/k_F)h/2e^2$. In order not to cause overfitting of the WL/WAL curves, it is essential to simplify Eq. (1) before fitting it to the experimental data.

The intrinsic KM SOC is a full invariant of the point group transformation of the crystal in question, which causes spin relaxation in graphene through the Elliott-Yafet (EY) mechanism and favors the existence of quantum spin Hall (QSH) state[26]. On the other hand, the extrinsic BR SOC produces spin relaxation in the graphene through the D'yakonov-Perel (DP) mechanism, which requires the breaking of the mirror symmetry of the graphene plane, therefore tends to destroy QSH state. In our experiment, hydrogen atoms are only bonded to one surface of graphene, which clearly breaks the mirror symmetry of graphene. Hence the KM term does not enter into the fitting. Furthermore, it has been shown that terms with the intravalley scattering rate usually provide small correction to the graphene WL signal[27, 28]; therefore, related terms can also be dropped. With the above considerations, Eq. (1) becomes:

$$\frac{2\pi h}{e^2}\frac{\Delta\rho}{\rho^2} = F\left(\frac{B}{B_\phi}\right) - F\left(\frac{B}{B_\phi+2B_i}\right) - 2F\left(\frac{B}{B_\phi+B_{BR}}\right) + 2F\left(\frac{B}{B_\phi+2B_i+B_{BR}}\right)$$
$$-F\left(\frac{B}{B_\phi+2B_{BR}}\right) + F\left(\frac{B}{B_\phi+2B_i+2B_{BR}}\right) \quad \text{Eq. (2)}$$

As shown in figure 2, Eq. (2) fits to the magneto-transport data very well and the resultant fitting parameters $B_x = \frac{\hbar}{4eL_x^2}$, $x = \phi, i, BR$, can be extracted. Among these parameters, $L_{BR}$ and $L_\phi$ are shown in figure 3 as a function of $n_H$ (Fig. 3 (a)&(b)), $V_g$ (Fig. 3 (c)&(d)) and temperature (Fig. 3 (e)&(f)). We shall discussion these two important parameters in detail in the following paragraphs.

First of all, the BR SOC scattering length $L_{BR}$ is found to reduce for higher $n_H$ (Fig. 3(a)), which strongly support our assertion that single-sided hydrogenation of graphene enhances its overall Rashba SOC strength. Second, $L_{BR}$ is also found to reduce as the Fermi level is getting close to the Dirac point (Fig 3(c)), which is consistent with other graphene hydrogenation experiment[3], except for the case of graphene/transition metal dichalcogenide heterostructure, where SOC strength is reported to slightly decrease towards the Dirac point[24, 29]. In addition, the gate tunability of SOC in hydrogenated graphene is much higher than that in graphene/TMD heterostructure[24, 29, 30], which is essential for manipulating and utilizing the spin degree of freedom in graphene-based device. It is worth noting that with increasing $n_H$, $L_{BR}$ and the momentum scattering length $L_{mfp}$ are both decreasing; $L_{BR}$ seems positively related to $L_{mfp}$ (see Supplementary Information S3). This appears to be in contrast to conventional expectation of the DP mechanism. We argue that such positive correlation between $L_{BR}$ and $L_{mfp}$ is due to the fact that *in-situ* hydrogenation is adding Rashba SOC at random sites of the crystal. Thus, while the addition of hydrogen increases the overall Rashba interaction and resulting a smaller $L_{BR}$, the resonance scattering from these hydrogen adatoms impurities[31] suppresses $L_{mfp}$. Two additional factors mentioned in above paragraphs support the DP mechanism: 1) hydrogenation at the top surface of the graphene clearly breaks the mirror symmetry; 2) KM SOC will not induce WAL observed in the hydrogenated samples.

Now we turn to the discussion of $L_\phi$, the phase coherence length. $L_\phi$ shows similar behavior to $L_{BR}$, which indicates that phase breaking scattering is enhanced as increasing $n_H$ (Fig 3(b)), decreasing $n_e$ (Fig 3(d)) or increasing temperature (Fig 3(f)), which agrees with other works on WL/WAL in graphene[24, 29, 32]. The phase breaking rate $\tau_\phi^{-1}$ of graphene is calculated from $\tau_\phi^{-1} = D/L_\phi$, where $D = v_F L_{mfp}/2$ is the diffusion constant. It is known that phase-breaking rate in graphene can be described as[20, 33]:

$$\tau_\phi^{-1} = aT + bT^2 + \tau_{sat}^{-1} \qquad \text{Eq. (3)}$$

The first term (linear in temperature $T$) in Eq. (3) is attributed mainly to the electron-electron interactions of 2D electron gas in the diffusive regime[20, 33-36], as well as the electron-phonon scattering process[35]; the second term ($\propto T^2$) in Eq. (3) in considered to arise from electron-electron interactions in graphene[20, 33]. At low temperatures, $\tau_\phi^{-1}$ should saturate to a constant term $\tau_{sat}^{-1}$ according to Eq. (3). Figure 4(a) plots the temperature dependence of $\tau_\phi^{-1}$ for graphene after the 9$^{th}$ round of hydrogenation and at different $V_g$ (different $n_e$). It is clear that $\tau_\phi^{-1}$ at all carrier density is decreasing at lower temperature and shows saturation behavior at low temperature.

The constant term $\tau_{sat}^{-1}$ in Eq. (3) for graphene with magnetic adatoms is found to be the spin-flip scattering rate $\tau_{sf}^{-1}$[20], which can be described by the Nagaoka-Suhl formula[19, 20, 37]:

$$\tau_{sf}^{-1} = \frac{n_{mag}}{\pi \hbar N(E_F)} \frac{\pi^2 S(S+1)}{\pi^2 S(S+1) + \ln^2(T/T_K)} \qquad \text{Eq. (4)}$$

where $n_{mag}$ is the density of magnetic impurities, $S$ is the total spin of the magnetic impurities, which is taken as 1/2 for hydrogen adatoms on graphene[38]. In addition, $T_K$ is the Kondo temperature, which is defined as $k_B T_K = A\exp[-1/N(E_F)J]$, with $A = 10$eV the cutoff energy, $J$ the Kondo exchange energy and $N(E_F) = 2\sqrt{n}/(\sqrt{\pi}\hbar v_F)$ the density of states at the Fermi level. Using the estimated hydrogen atom density[13] as $n_{mag}$, $T_K$ is found to be much less than 1K, which is consistent with our experimental observation. In the meantime, due to the small $T_K$, $\tau_{sf}^{-1}$ does not change much in the temperature range of this experiment (5.8 K – 60 K). The $n_H$ and $n_e$ dependence of $\tau_{sat}^{-1}$ is plotted in figure 4(b) and 4(c), respectively. $\tau_{sat}^{-1}$ increases with more hydrogen atoms absorbed on the graphene surface and decreases when the Fermi surface of the graphene moves away from the Dirac point, which is consistent with the expectation of the Nagaoka-Suhl formula. This is an important evidence showing that hydrogen adatoms on graphene are indeed magnetic and can induce spin-flip scattering.

### III. CONCLUSIOIN

In summary, we have investigated low field magneto-transport properties of *in-situ* hydrogenated graphene and observed hydrogen density and carrier density tunable WL to WAL transition. By analyzing the low field magnetoresistance, we determined the strength of Bychkov-Rashba type spin-orbit coupling can be efficiently tuned with hydrogen density and carrier density. From the temperature dependent phase breaking scattering rate, the saturation scattering rate is deduced and confirmed to be spin-flip scattering, which is an important evidence for magnetic scattering induced by hydrogen adatoms on graphene.

### IV. ACKNOWLEDGEMENT

The authors thank X. C. Xie and S. Q. Shen for helpful discussion and C. Y. Cai for help on experiment. This project has been supported by the National Basic Research Program of China (Grant Nos. 2019YFA0308402，2018YFA0305604), the National Natural Science Foundation of China (NSFC Grant Nos. 11934001，11774010，11921005), Beijing Municipal Natural Science Foundation (Grant No. JQ20002).

**Figure 1**

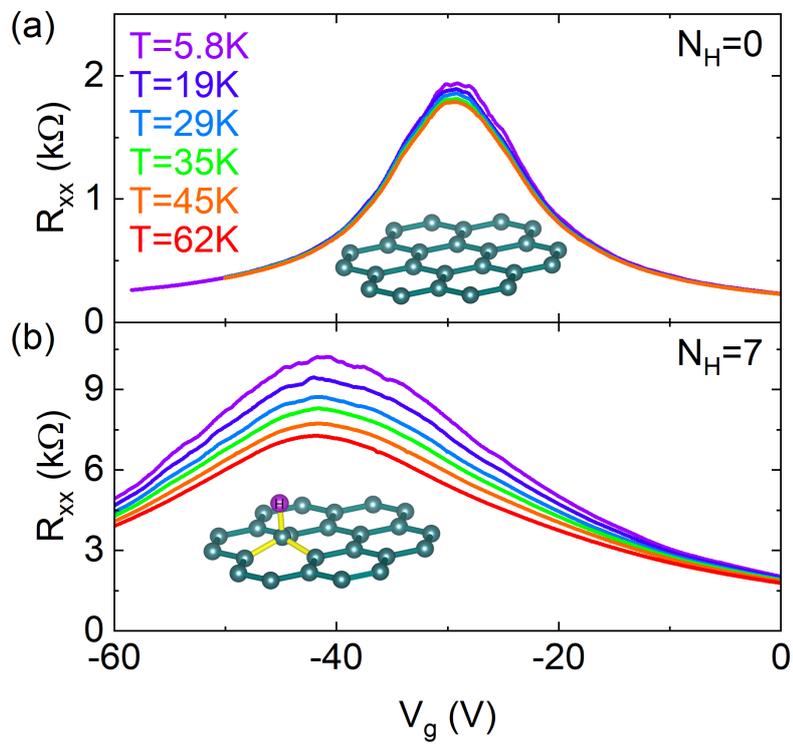

Figure 1. Resistance as a function of back gate voltage $V_g$ of (a) pristine and (b) hydrogenated graphene, respectively. $N_H$ denotes hydrogenation rounds, e.g. $N_H=7$ means after 7th round of hydrogenation process. Insets in the two panels are illustrations for the atomic structure of un-doped and hydrogen-doped graphene.

**Figure 2**

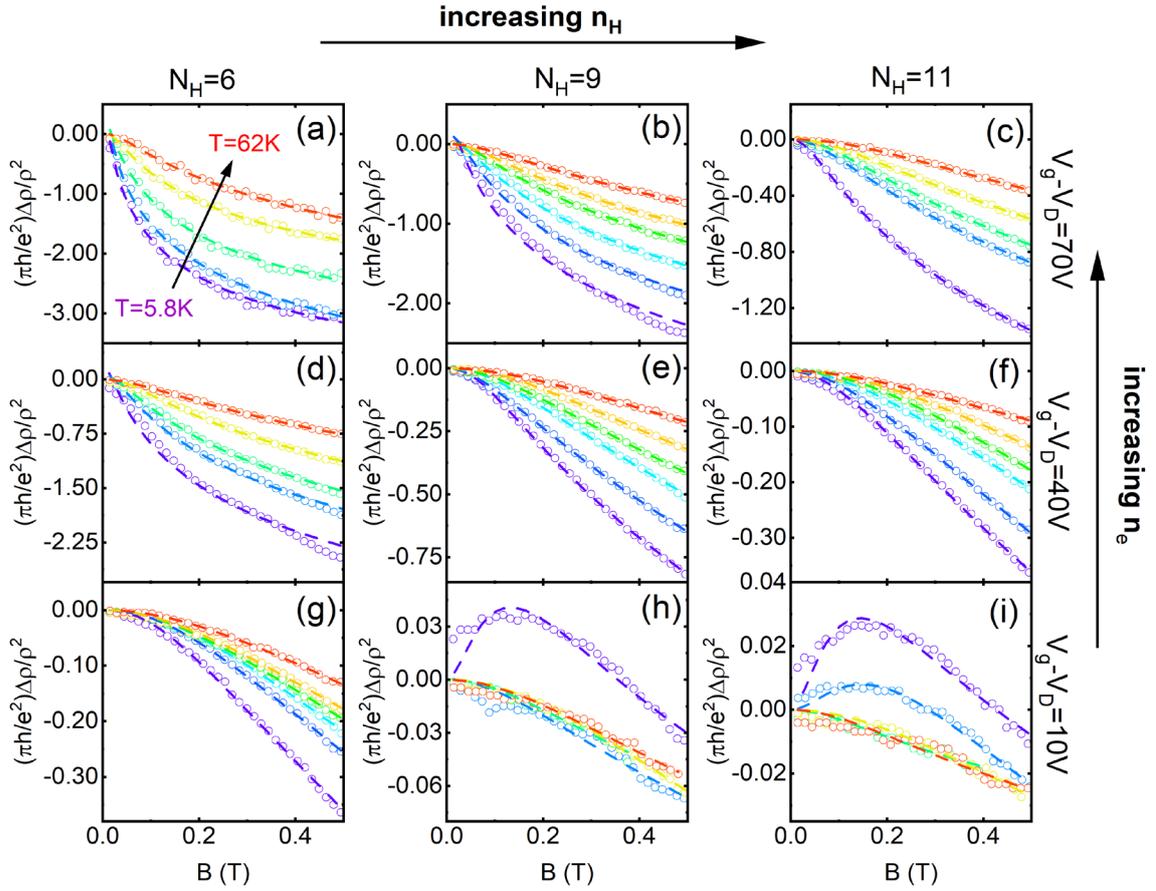

Figure 2. Low field (B < 0.5 T) magnetoresistance (*MR*) of hydrogenated graphene at various temperature *T*, carrier density $n_e$ and hydrogen density $n_H$. *MR* behavior of hydrogenated graphene shows transition from WL to WAL at high $n_H$ and low $n_e$ (e.g., $V_g$-$V_D$≤10V); WL suppression is observed at high $n_H$ and high $n_e$ (e.g., $V_g$-$V_D$≥40V). *MR* of up to 9 T magnetic field can be found in Ref[13]. The broken lines in the panels are fit to Eq.(2).

**Figure 3**

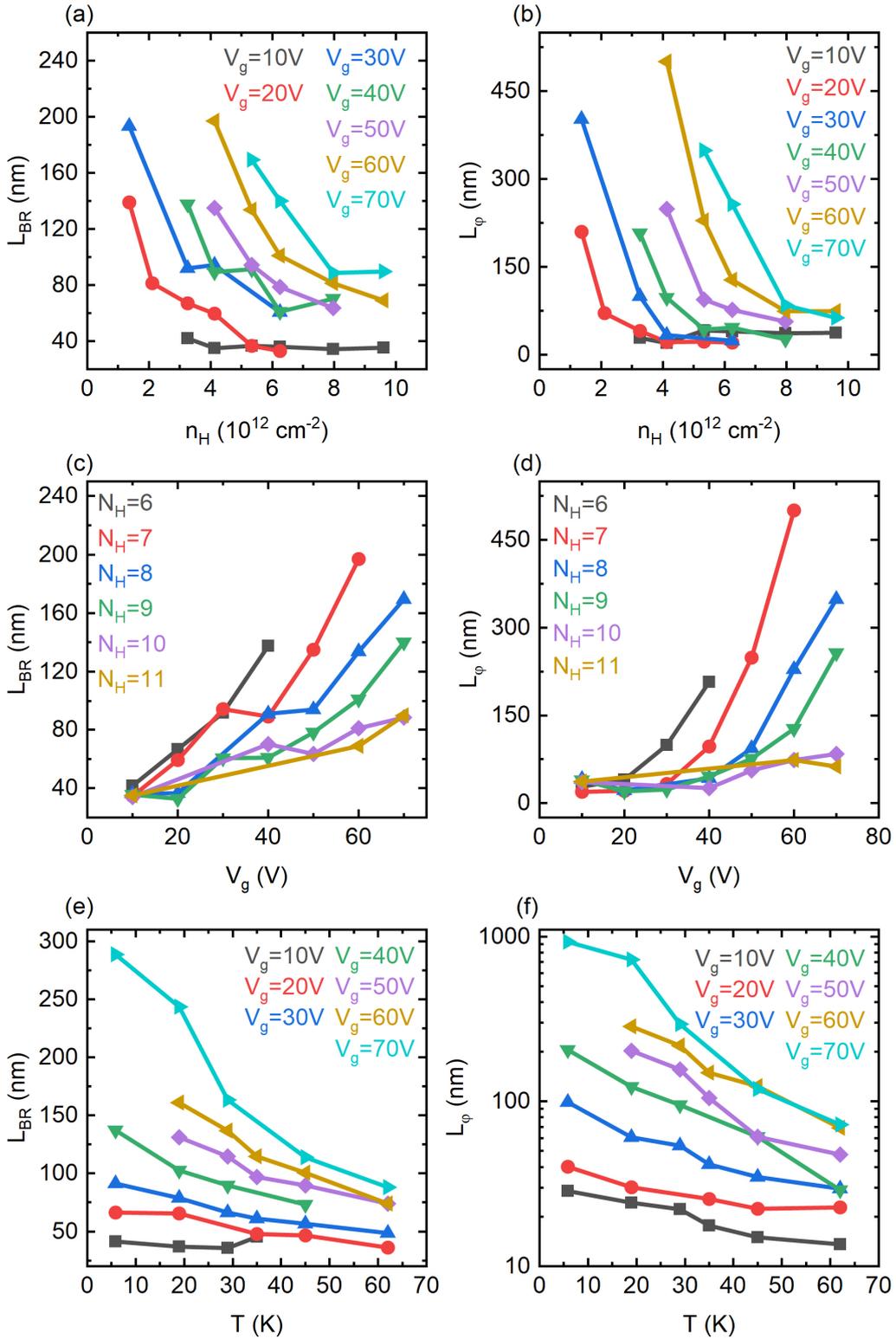

Figure 3. WAL fitting results of hydrogenated graphene. (a),(c),(e) The BR SOC scattering length $L_{BR}$ vs. $n_H$, $V_g$ and $T$, respectively. (b),(d),(f) The phase breaking scattering length $L_\varphi$ vs. $n_H$, $V_g$ and $T$, respectively. (e) and (f) are taken from $N_H=6$ data.

**Figure 4**

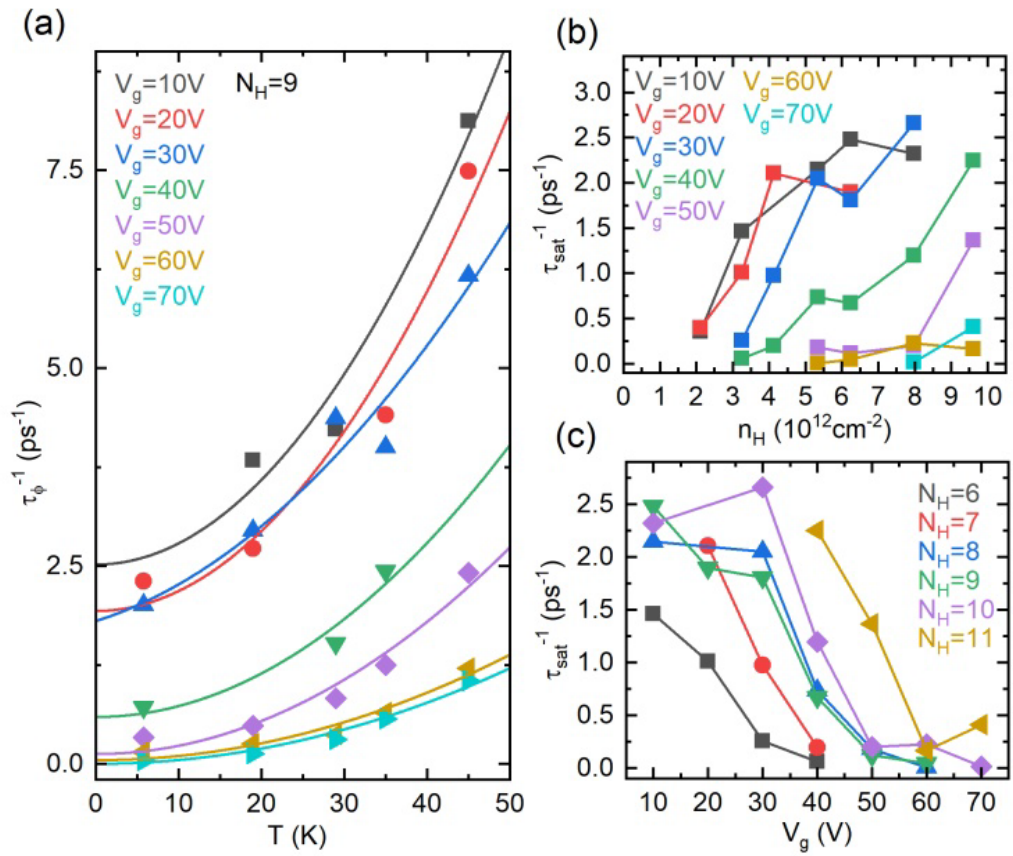

Figure 4. *T* dependence and saturation of phase breaking rate $\tau_\phi^{-1}$. (a) Experimental $\tau_\phi^{-1}$ vs. *T* data (dots) and fitting to Eq. (3) (solid line). (b)&(c) Saturated phase breaking rate $\tau_{sat}^{-1}$ vs. $n_H$ and $V_g$, respectively.